\documentclass{ws-procs975x65}

\def\beq{\begin{equation}}
\def\eeq{\end{equation}}

\newcommand{\mb}[1]{\mathbf{#1}}

\begin{document}

\title{Thermodynamic Analysis of a Self-Gravitating Gas in 
Astrophysical Contexts}
\author{Christine Gruber$^*$}

\address{Hanse-Wissenschaftskolleg, Delmenhorst, Germany \\
and \\
Institute of Physics, Carl v. Ossietzky University of Oldenburg, Germany\\
$^*$E-mail: christine.gruber@uni-oldenburg.de}

\begin{abstract}
The thermodynamics of a self-gravitating gas cloud of particles 
interacting only via their gravitational potential is an interesting 
problem with peculiarities arising due to the long-ranged nature of 
the gravitational interaction. Based on our recent work on the 
properties of such a configuration, we extend the system to contain 
a central gravitational field in which the particles are moving, to 
mimic the potential of a central compact object exerting an external 
force on the gas cloud. After an introduction to the general problem, 
including the aforementioned peculiarities and possible solutions, 
we will discuss the particular properties of the self-gravitating gas 
in a central field and its thermodynamic analysis. 
\end{abstract}


\bodymatter

\section{Introduction}
The topic of thermodynamic systems in the presence of gravity has 
been discussed in many occasions and forms 
\cite{1968Lynd1,2003Oppenheim1,2009Campa1}, and has raised 
many questions on how to deal with the long-range effects of gravity 
in the thermodynamic analysis of systems, where concepts such as the 
isolation of a system in one or the other regard are important. 
Non-stationary equilibrium situations, negative heat capacities or 
simply divergences in the thermodynamic limit have been plaguing the 
analyses, and the conventional, very successful thermodynamic framework 
of Boltzmann-Gibbs statistics had to be adapted and modified in order 
to account for the peculiarities of the thermodynamics of a gravitational 
system. 

Based on a Boltzmann-Gibbs analysis of the self-gravitating gas 
\cite{2002VS1}, the statistical analysis and subsequent calculation of 
thermodynamic properties have been carried out \cite{2018Esca1} 
assuming a generalized framework  intended to describe a system 
with non-extensive properties, due to the presence of long-range 
forces such as gravity. The adopted generalized framework, i.e., 
Tsallis generalized $q$-statistics, has been developed in order to 
consider non-extensive effects, entailing an additional parameter 
$q$ in the statistical analysis. 

This work is an extension of these previous investigations which 
generalizes the self-gravitating gas to a more realistic system 
featuring a centrally placed compact object, like e.g. a black hole, 
around which the gas is extending. Due to some peculiarities and open 
questions, we will not continue using the non-extensive $q$-statistics, 
but rather return to the conventional Boltzmann-Gibbs statistics, in 
order to get a first impression of the results. Other generalizations 
can be thought of, which will be commented on in the last section.

\section{Statistical mechanics and thermodynamics of a self-gravitating gas}
I will briefly review the most important steps in the analysis 
of a self-gravitating gas \cite{2018Esca1,2002VS1}, from the system's 
properties to the peculiarities of the thermodynamic analysis and 
some of its outcomes. 

The governing force of the self-gravitating gas is the gravitational 
attraction between its $N$ identical constituent particles which are 
otherwise moving freely, and thus the Hamiltonian of the system is 
\begin{equation} \label{eq:H}
	\mathcal{H} = \mathcal{T} + \mathcal{U} 
		= \sum_{i=1}^{N}\frac{p_i^2}{2m}-Gm^2 
		\sum_{1\leq i<j \leq N} 
		\frac{1}{\left| \mb{q_i}-\mb{q_j} \right|_A} \,,
\end{equation}
where $G$ is the gravitational constant, $m$ the mass of an individual 
particle, and $A$ represents a short-range cutoff imposed in order to 
avoid the unphysical collapse of the system to a point. 
This Hamiltonian is the basis for a thermodynamic analysis which 
can be done in principle in different ensembles, like the 
microcanonical one, where the energy of the system is kept constant, 
or the canonical one, where instead the temperature is fixed, and 
energy can be exchanged with a reservoir. In the microcanonical 
ensemble, the most important thermodynamic quantity from which 
everything is derived is the entropy, i.e., the logarithm 
of this sum over microstates $\Omega (E,V,N)$, 
\begin{equation}\label{eq:S-MCE}
	S = k_B \ln \Omega (E,V,N) \,,
\end{equation}
where 
\begin{equation}\label{eq:Omega}
	\Omega = \frac{(2\pi m)^{3N/2}}{N! \,h^{3N}
	\Gamma\left(\frac{3N}{2}+1\right)}
	\int d^{3N}q\ \big[E-\mathcal{U}\big]^{3N/2-1} \,. 
\end{equation}
From the entropy, you can obtain important thermodynamic quantities 
such as the temperature of the gas, or the equation of state, i.e., 
the relation between pressure, temperature and volume. \\
In the case of the canonical ensemble, the starting point is the 
partition function, defined as 
\begin{equation}
	Z = \frac{1}{N!h^{3N}}\int d^{3N}q \, d^{3N}p \, 
	\exp_q{\left(-\beta \mathcal{H}(\textbf{p},\textbf{q})
	\right)} \,,
\end{equation}
and everything else is derived from that quantity, like the equation 
of state. The temperature in this ensemble is fixed, so it cannot be 
calculated. \\
Following the definition of these basic thermodynamic functions, 
calculations can be simplified by the assumption of a weak 
gravitational interaction, i.e., the gravitational contribution can 
be treated as a small correction to the ideal gas, and results can be 
obtained analytically in this case. \\
Under this additional assumption, further quantities that are of 
thermodynamic interest can be calculated, like the heat capacity and 
other response functions of the system. Both equation of state and heat 
capacity have been calculated and compared in the framework of Tsallis 
statistics \cite{2018Esca1}, and the details can be found there. \\
An important point for further investigations is the question of the 
statistical framework, which is closely connected to the choice of 
thermodynamic limit. Tsallis' non-extensive statistics naturally 
features a modification of the thermodynamic limit, in which the 
thermodynamic state variables result in convergent functions. In 
the case the conventional Boltzmann-Gibbs statistics, another 
modification of the thermodynamic limit has to be adopted \cite{2002VS1} 
in order to obtain convergent results. In the following, we will 
employ Boltzmann-Gibbs statistics with the modified thermodynamic 
limit.

\section{Addition of the central gravitational potential}
As a modification to the basic setup of a simple self-gravitating 
gas many complications can be thought of. The simplest gase perhaps 
is the addition of a central potential, to model the situation of a 
self-gravitating gas around a black hole. We will start with the 
assumption of an external gravitational field caused by a mass $M$ 
of size $r_S=2GM/c^2$ in the center of the configuration, restricting 
the movement of the gas between the radius of the innermost stable 
circular orbit (ISCO) at $r_{ISCO} = 3 r_S$ and infinity. This will 
make a difference in the integrals contained in the sum over 
microstates and the partition function, respectively. Moreover, the 
central potential will have its influence on every particle in the gas. 
The generalized Hamiltonian thus reads 
\begin{equation} \label{eq:Hgen}
	\mathcal{H} = \sum_{i=1}^{N}\frac{p_i^2}{2m}-Gm^2 
	\sum_{1\leq i<j \leq N} \frac{1}{\left| \mb{q_i}-\mb{q_j} 
	\right|_{A}} - G m M \sum_{1\leq i<j \leq N} \frac{1}{\left| 
	\mb{q_i}-\mb{r} \right |_{A}} \,,
\end{equation}
with $\mb{r}$ denoting the center of mass of the compact object. To 
simplify calculations, we choose $\mb{r}=0$. \\
The computational procedures in order to extract the thermodynamic 
equation of state is fairly analogous to the case of a simple 
self-gravitating gas, and differs only in the restriction of the 
range of integration, due to the fact that we consider a ring-like 
structure, or even a flat two-dimensional disk shape. This restriction 
will manifest itself in the definition of the virial coefficients 
$b_i$, which will be slightly different. \\
The interesting question is whether the modification of the system 
will lead to differences in the thermodynamic limit, i.e., facilitate 
the calculation of otherwise divergent functions, or modify the 
qualitative dependence on the number of particles in the thermodynamic 
limit. Preliminary results indicate that this is not the case, and 
that modifications are limited to the virial coefficients of the 
problem.

\section{Outlook}
We have here discussed the thermodynamic properties of a 
self-gravitating gas under the influence of a central gravitational 
field caused by a heavy mass at the center of the configuration. 
Basing on the analysis of a self-gravitating gas cloud consisting 
of ideal particles \cite{2018Esca1,2002VS1}, an additional term 
accounting for the central gravitational potential was added to 
the analysis, and the resulting thermodynamic state variables were 
calculated. Preliminary results indicate slight modifications of 
the state variables, depending on the new parameter, the mass of 
the central object. The goal is to generalize the analysis of a 
simple self-gravitating gas to eventually be able to make 
predictions on the thermodynamic behavior of matter around a 
compact object, i.e., an accretion disk of sorts. 

Besides the inclusion of a central compact object, the gas itself 
can be modified in its properties, e.g., by considering non-ideal 
interactions between the particles. This could be accounted for in 
an exact way by adding additional particle-particle interactions to 
the Hamiltonian, with the corresponding coupling constant, like for 
example an electromagnetic charge. The different strengths of gravity 
and any other interactions that may be added have to be weighed 
against each other, and approximations could be applied. Another 
possibility would be to include effective potentials which are used 
in condensed matter systems, Mie-type potential like the Lennard-Jones 
case or others, in order to describe different variations of the gas. 
Investigations in this direction would represent the first steps 
towards the description of non-ideal fluids in gravitational contexts - 
either gas clouds of interacting particles, or non-ideal fluids 
constituting accretion disks or clouds around a central compact 
object. 

Further generalizations include rotation of the system, or charge 
of the central object. 
Importantly, these results should then be connected to results of 
other calculations on accretion disks, in particular the accretion 
of (charged) dust particles in a spherical shell or torus structure 
\cite{2017Schr1}.

\section*{Acknowledgments}
C. G. was supported by funding from the DFG 
Research Training Group 1620 `Models of Gravity'.

\bibliographystyle{ws-procs975x65}
\bibliography{BHTD}

\end{document}